\documentclass[letterpaper]{article} % DO NOT CHANGE THIS
\usepackage{aaai24}  % DO NOT CHANGE THIS
\usepackage{times}  % DO NOT CHANGE THIS
\usepackage{helvet}  % DO NOT CHANGE THIS
\usepackage{courier}  % DO NOT CHANGE THIS
\usepackage[hyphens]{url}  % DO NOT CHANGE THIS
\usepackage{graphicx} % DO NOT CHANGE THIS
\usepackage{natbib}  % DO NOT CHANGE THIS AND DO NOT ADD ANY OPTIONS TO IT
\usepackage{caption} % DO NOT CHANGE THIS AND DO NOT ADD ANY OPTIONS TO IT
\usepackage{amsmath}
\usepackage{amssymb}
\newtheorem{Definition}{Definition}
\newtheorem{Assumption}{Assumption}

\newtheorem{Theorem}{Theorem}

\newtheorem{Lemma}{Lemma}

\DeclareMathOperator*{\argmax}{argmax}

\DeclareMathOperator*{\argsup}{argsup}
\usepackage{algorithm}
\usepackage{algorithmic}

\usepackage{newfloat}
\usepackage{listings}
\DeclareCaptionStyle{ruled}{labelfont=normalfont,labelsep=colon,strut=off} % DO NOT CHANGE THIS
\floatstyle{ruled}
\newfloat{listing}{tb}{lst}{}
\floatname{listing}{Listing}
\title{Optimal Mechanism in a Dynamic Stochastic Knapsack Environment}
\author{
    Jihyeok Jung\textsuperscript{\rm 1}, 
    Chan-Oi Song\textsuperscript{\rm 2}, 
    Deok-Joo Lee\textsuperscript{\rm 1}\thanks{Deok-Joo Lee is the corresponding author.}, 
    Kiho Yoon\textsuperscript{\rm 2}}
\affiliations{
    \textsuperscript{\rm 1}Department of Industrial Engineering, Seoul National University \\
    \textsuperscript{\rm 2}Department of Economics, Korea University\\
    firedaeman@snu.ac.kr, 
    gnos@korea.ac.kr,
    leedj@snu.ac.kr,
    kiho@korea.ac.kr
}

\begin{document}

\maketitle

\begin{abstract}
This study introduces an optimal mechanism in a dynamic stochastic knapsack environment. The model features a single seller who has a fixed quantity of a perfectly divisible item. Impatient buyers with a piece-wise linear utility function arrive randomly and they report the two-dimensional private information: marginal value and demanded quantity. We derive a revenue-maximizing dynamic mechanism in a finite discrete time framework that satisfies incentive compatibility, individual rationality, and feasibility conditions. It is achieved by characterizing buyers' utility and deriving the Bellman equation. Moreover, we propose the essential penalty scheme for incentive compatibility, as well as the allocation and payment policies. Lastly, we propose algorithms to approximate the optimal policy, based on the Monte Carlo simulation-based regression method and reinforcement learning.
\end{abstract}

\section{Introduction}
Dynamic resource allocation refers to the distribution of a limited amount of resources in a dynamic environment. This problem arises in various fields where the system has a fixed capacity to serve time-varying demands, such as cloud computing and software-as-a-service \citep{Wu2011, Wang2013, zhang2015}. Therefore, it is crucial for system designers to find an optimal policy to achieve their desired objectives while considering the resource provision for future demands. Although the structure of dynamic allocation problems can vary depending on the objectives and assumptions, this study focuses on a specific category of dynamic resource allocation, the \textit{Dynamic Stochastic Knapsack Problem} (DSKP).

In the original DSKP, the seller tries to sell a fixed amount of an item over a finite time horizon. During each time period, customers with information about the item's value and desired quantity enter the system. The seller, with stochastic knowledge of customer arrivals, aims to determine the optimal allocation strategy to maximize the total expected value \citep{Papastavrou1996, Kleywegt1998}. Moreover, the original DSKP assumes that arriving demands are non-strategic. As each buyer is assumed to behave non-strategically, the problem of maximizing expected value naturally aligns with the goal of maximizing expected revenue, since the value can be interpreted as the maximum willingness to pay.

However, this assumption about non-strategic buyers does not hold true in many real-life scenarios. In reality, customers might strategically misreport their information to achieve favorable outcomes. They could request quantities exceeding their actual desire or overbid the resource's value in an attempt to improve their chances of securing an allocation priority. Given this potential, the seller needs to devise an optimal selling mechanism that maximizes expected revenue when buyers are behaving strategically.

If all participants act strategically, the aforementioned problem can be effectively tackled through mechanism design. By extending the work of \citet{Myerson1981}, which proposed the optimal mechanism in one-dimensional and static environments, we can derive the optimal dynamic mechanism in two-dimensional and dynamic environments. However, it's important to note that dynamic mechanisms often involve intricate mathematical formulations, which can pose challenges when directly applied to real-world service systems. To overcome these complexities, the development of approximation algorithms for the derived mathematical solutions becomes essential. This approach aims to render these mechanisms practically implementable, mitigating the challenges associated with their mathematical complexity.

To this end, we consider a revenue-maximizing dynamic mechanism under the dynamic stochastic knapsack environment. Buyers arrive randomly throughout the finite time horizon with private two-dimensional information about their desired quantities and values, which follows a continuous joint probability distribution. In this setting, the optimal allocation and payment rules that satisfy incentive compatibility and individual rationality are derived by modifying the characterization approach and dynamic program. Furthermore, we propose two algorithms to approximate the proposed mathematical solution by implementing Monte Carlo (MC) simulation-based regression method and reinforcement learning. We compare and analyze the performance of these algorithms to evaluate their effectiveness.

The suggested model and algorithms in this study contribute in the following ways:
\begin{itemize}
    \item We propose a general structure for an optimal dynamic mechanism by relaxing assumptions presented in existing two-dimensional dynamic mechanisms. Firstly, we assume that buyers have a piece-wise linear function instead of the conventional take-it or leave-it form. Secondly, we extend the buyers' type space and the seller's decision set as continuous domains. Lastly, we address an environment where the number of arriving buyers at each time period remains unknown.
    \item We determine allocation and payment policies contingent on the time period, remaining item quantity, and submitted bids from incoming buyers. Moreover, we devise a penalty scheme to ensure incentive compatibility, which is necessary to prevent buyers from overbidding for the demanded quantity. 
    \item We introduce two algorithms to numerically approximate the proposed mechanism. The first algorithm employs a Monte Carlo (MC) simulation-based regression method, which approximates the state value function with a polynomial function. The second algorithm relies on the deep deterministic policy gradient (DDPG) technique, focusing on learning the allocation policy.
\end{itemize}

\section{Related Works}
\subsection{Dynamic Stochastic Knapsack Problem}
DSKP emerged as a component of the dynamic allocation problem and was initially analyzed in a finite discrete-time framework by \citet{Papastavrou1996}. Subsequently, \citet{Kleywegt1998} extended this framework to encompass continuous-time scenarios, while \citet{Kleywegt2001} investigated with random-sized demands. They aimed to find optimal allocation and pricing rules within a non-strategic demand context. Notably, they suggested that the optimal allocation and pricing rules are threshold policies, which offer a uniform price to incoming buyers and the buyers either accept or reject the offer.

Furthermore, various computation algorithms for DSKP have been developed to implement allocation policies in practical scenarios \citep{Zhou2008, han2015randomized, Im2021, Sun2022}. They focused on online algorithms to allocate resources in response to changing demands and constraints. They offered theoretical bounds to guarantee the efficiency of the proposed algorithms. However, the presence of strategic behaviors among buyers necessitates a mechanism design approach to maximize the seller's expected revenue.

\subsection{Optimal Mechanism}
In the field of mechanism design, the solution to the presented problem—an optimal mechanism in a dynamic stochastic knapsack environment—can be viewed as a two-dimensional type space (value and quantity), multi-unit (perfectly divisible item) dynamic mechanism. Our approach effectively adapts the findings of mechanisms in a static environment to span multiple periods. The cornerstone of the optimal mechanism in a static environment is the work of \citet{Myerson1981}, which revolved around a revenue-maximizing auction for a single item using the characterization approach.

Subsequently, much of the literature has expanded to encompass various settings. \citet{Che1993} and \citet{Asker2010} explored the multi-dimensional bidding environment, while \citet{Maskin1989} extended the model to a multi-unit setting. Furthermore, \citet{Iyengar2008} formulated a two-dimensional and multi-unit procurement auction mechanism with perfectly divisible items, laying the groundwork for constructing a periodic allocation model in our study. Additionally, to address the computational challenges of the optimal mechanism, \citet{Bhat2019} suggested an algorithm that transforms integration into a summation operation, and \citet{Dutting2019} used a neural network to approximate the optimal mechanism in various auction environments.

\subsection{Dynamic Mechanism}
In the context of the dynamic mechanism based on the Myersonian approach, \citet{Bergemann2019} noted that the optimal dynamic mechanism has been explored in two main strands. The first strand involves fixed participating agents with evolving values over time \citep{kakade2013optimal, pavan2014dynamic, pavan_2017}. However, our approach diverges from these works as we consider a scenario where the buyers are impatient, meaning they stay in the system only temporarily. 

Our study aligns with the second strand, where the population of bidders changes over time while retaining fixed private values. This line of research primarily focuses on revenue management for sellers dealing with the requests of impatient buyers. \citet{Vulcano2002} tackled the challenge of designing an optimal mechanism for selling indivisible, limited items in a discrete-time space, involving randomly arriving unit-demand buyers. Building upon this, \citet{Gershkov2009} extended the problem to a scenario with heterogeneous goods and continuous-time arrivals of buyers and \citet{pai2013optimal} considered bids in three dimensions (value, arrival time, deadline) from unit-demand buyers.

The two-dimensional optimal dynamic mechanism where the buyers bid their value and quantity was studied in \citet{Dizdar2011} and \citet{Wang2013}. The former considered an environment where a single buyer arrives every period, while the latter assumed the items are indivisible discrete units. Additionally, both studies simplified the analysis by considering the buyer's utility function as a take-it or leave-it form, where the utility becomes zero if the desired quantity is not obtained. In our model, we aim to improve upon these aspects.

\section{Framework}

A monopolistic seller possesses $\bar{Q}$ units of a perfectly divisible item. The seller plans to sell the item during a finite discrete time horizon $\mathcal{T} = \{1, \cdots, T\}$ and aims to maximize the ex-ante expected revenue at the beginning. In every period $t \in \mathcal{T}$, $n^t$ buyers arrive in the market, where $n^t$ follows a probability mass function $g(n)$ with the support $\{1, \cdots, N\}$. Every buyer is assumed to be \textit{impatient}: the buyer arriving at $t$ leaves the market before $t+1$.

The arriving buyer at time $t$ is denoted as $(i,t)$, where $1 \le i \le n^t$. The buyer $(i,t)$ has private information $\theta^t_i = (v_i^t, q_i^t)$, with $v_i^t$ representing the marginal value of the item and $q_i^t$ representing the desired quantity. $\theta_i^t$ is independently distributed among the $n^t$ bidders and follows a joint probability density function $f(v, q)$ and cumulative distribution function $F(v, q)$ with the support (or type space) $\Theta = [\underline{v}, \bar{v}] \times [\underline{q}, \bar{q}]$. The lowest values are normalized to 0, i.e., $\underline{v} = 0$ and $\underline{q} = 0$.

We define $\theta^t = (\theta^t_1, \cdots, \theta^t_{n^t}, \varnothing, \cdots, \varnothing) \in \Theta^{N}$ as the type profile of bidders at period $t$. The former part describes the type vector of arriving buyers, and the latter describes the dummy types of non-arriving buyers. Here, we let $\varnothing = (0, 0)$ to ensure that the dummies do not influence the allocation results. By defining $\theta^t$ with dummies, we can express the allocation and payment rules as $N$-dimensional functions. Additionally, we define $\theta^t_{-i}$ as the type profile excluding buyer $(i,t)$.

Then, by the revelation principle, we can focus on the incentive-compatible direct mechanism where buyers truthfully report their types.

\begin{Definition}\normalfont \textbf{(Direct Mechanism)}
\textit{$\Gamma = (a^t, p^t)_{t \in \mathcal{T}}$ is a direct mechanism where
\begin{equation}
a^t: \Theta^N \times [0, \bar{Q}] \rightarrow \mathbb{R}_{+}^{N} \mbox{ and } p^t: \Theta^N \times [0, \bar{Q}] \rightarrow \mathbb{R}_{+}^{N} \nonumber
\end{equation}
are the allocation rule and payment rule at period $t$. }
\end{Definition}

The direct mechanism comprises a sequence of time-dependent allocation and payment rules. They are determined based on the reported type profile and the remaining units of the item. Let $q^t$ be the remaining units at the beginning of period $t$, which is public information. Then, the allocation and payment for buyer $(i,t)$, who reports their type as $\hat{\theta}_i^t = (\hat{v}_i^t, \hat{q}_i^t)$ when others bid truthfully, are denoted as $a_i^t(\hat{\theta}^t_i, \theta^t_{-i}, q^t)$ and $p_i^t(\hat{\theta}^t_i, \theta^t_{-i}, q^t)$, respectively. Based on this, we define the utility function for the buyers as follows.

\begin{Definition}\normalfont \textbf{(Ex-Post Utility Function)}
\textit{
Given $q^t$ units remaining in the market at time $t$, the ex-post utility of buyer $(i,t)$, who reports $\hat{\theta}_i^t$ with the true type $\theta^t_i$, is defined as:
\begin{multline}
    u_i^t(\hat{\theta}^t_i, \theta^t_{-i} \vert \theta_i^t, q^t)  = v_i^t \min\{q_i^t, a_i^t(\hat{\theta}^t_i, \theta^t_{-i}, q^t)  \} \\ 
    - p_i^t(\hat{\theta}^t_i, \theta^t_{-i},  q^t).
\end{multline}}
\end{Definition}
The utility function for the buyers is piece-wise linear. The marginal utility remains constant at $v_i^t$ until the allocation reaches the demanded quantity $q_i^t$, and once it exceeds, the utility no longer increases. This utility function is a natural extension of the linear utility introduced by \citet{Myerson1981}. Subsequently, we define the following expectations for the Bayesian equilibrium.

\begin{Definition}\normalfont \textbf{(Expected Allocation and Payment)} \textit{Given $q^t$ units remaining in the market at time $t$, the expected allocation and payment of bidder $(i,t)$ who reports $\hat{\theta}_i^t$ with the true type $\theta^t_i$ is defined by
\begin{align*}
    & A_i^t(\hat{\theta}_i^t \vert \theta_i^t,  q^t) = \mathbb{E}_{\theta^t_{-i}}[ a_i^t(\hat{\theta}^t_i, \theta^t_{-i} , q^t)] \mbox{ and } \\
    & P_i^t(\hat{\theta}_i^t \vert \theta_i^t,  q^t) =  \mathbb{E}_{\theta^t_{-i}} [ p_i^t(\hat{\theta}^t_i, \theta^t_{-i}, q^t)],
\end{align*} 
provided that the other buyers report their types truthfully.} 
\end{Definition}

\begin{Definition}\normalfont\textbf{(Interim Utility Function)} \textit{
Given $q^t$ units remaining in the market at time $t$, the expected (interim) utility of bidder $(i,t)$ who reports $\hat{\theta}_i^t$ with the true type $\theta^t_i$ is defined by
\begin{equation*}
    U_i^t(\hat{\theta}_i^t \vert \theta_i^t, q^t) = \mathbb{E}_{\theta^t_{-i}}[ u_i^t(\hat{\theta}^t_i, \theta^t_{-i} \vert \theta_i^t,   q^t)], 
\end{equation*} 
provided that the other buyers report their types truthfully.}
\end{Definition}
For the sake of simplicity, denote the expected allocation, payment, and utility function when the buyer truthfully reports its type as $A_i^t(\theta_i^t \vert  q^t) = A_i^t(\theta_i^t \vert \theta_i^t,  q^t)$, $P_i^t(\theta_i^t \vert q^t) = P_i^t(\theta_i^t \vert \theta_i^t, q^t)$ and $U_i^t(\theta_i^t \vert  q^t) = U_i^t(\theta_i^t \vert \theta_i^t, q^t)$.

Then, the objective of the seller is to maximize the ex-ante expected revenue which can be written as
\begin{equation}
    \underset{a, p}{\max} \quad \sum_{t=1}^{T}{\delta^{t-1} \mathbb{E}_{n^{t}, \theta^{t}}  \left[\sum_{i=1}^{n^t}{p_{i}^{t}(\theta^{t},  q^t) }\right]},
\end{equation}
with the following constraints:
\begin{itemize}
    \item \textbf{Bayesian incentive compatibility}: For any $t \in \mathcal{T}$ and $0 \le q^t \le \bar{Q}$, every arriving buyer $(i,t)$ has no incentive to misreport its true type:
\begin{equation}
    U_i^t(\theta^t_i \vert q^t) \ge U_i^t(\hat{\theta}^t_i \vert \theta^t_i, q^t), \quad \forall \theta_i^t, \hat{\theta}^t_i \in \Theta.
\end{equation}
    \item \textbf{Individual rationality}: For any $t \in \mathcal{T}$ and $0 \le q^t \le \bar{Q}$, every arriving buyer $(i,t)$ should not be worse off by participating in the mechanism:
\begin{equation}
      U_i^t(\theta_i^t \vert q^t) \ge 0 \quad \forall \theta_i^t \in \Theta.
\end{equation}
    \item \textbf{Feasibility}: For any $t \in \mathcal{T}$, if the buyers report $\hat{\theta}^t$ and the seller has $q^t$ remaining units,
\begin{align}
 & \sum_{i=1}^{N}a_i^t(\hat{\theta}^t, q^t) \le q^t, \ q^{t+1} = q^t - \sum_{i=1}^{N}a_i^t(\hat{\theta}^t, q^t) \\
 & 0 \le a_i^t(\hat{\theta}^t, q^t) \le \hat{q}^t_i.
\end{align}
\end{itemize}
The feasibility condition (5) implies that the periodic allocation cannot exceed the current remaining units and there are no newly added items or returns. 
Also, $q^1 = \bar{Q}$.
Meanwhile, condition (6) implies the individual allocation cannot exceed the reported quantity.

\section{Optimal Mechanism}
In this section, we derive the optimal solution to the proposed problem. To achieve this, we initially conducted the characterization of the incentive-compatible mechanism. The proofs of the propositions marked with $\clubsuit$ are omitted here and can be found in the full version of the paper.
\begin{Lemma} \normalfont($\clubsuit$)
    \textit{Suppose $\Gamma=(a^t,p^t)_{t \in \mathcal{T}}$ is incentive compatible and feasible. Then at any period $t \in \mathcal{T}$, }
\begin{itemize}
\item[(a)] \textit{$U_{i}^{t}(v_{i}^{t},q_{i}^{t} \vert q^t)$ is convex with respect to $v_{i}^{t}$}.
\item[(b)] $\forall \varepsilon>0$, $U_{i}^{t}(v_{i}^{t},q_{i}^{t} \vert q^t)- U_{i}^{t}(v_{i}^{t} - \varepsilon,q_{i}^{t} \vert q^t) \leq \varepsilon A_{i}^{t}(v_{i}^{t}, q_{i}^{t} \vert q^t)$ $\leq  U_{i}^{t}(v_{i}^{t}+ \varepsilon, q_{i}^{t} \vert q^t) - U_{i}^{t}(v_{i}^{t},q_{i}^{t}\vert q^t)$.
\end{itemize}
\end{Lemma}
Given (a) of Lemma 1, $U_i^t(v_i^t, q_i^t \vert q^t)$ is absolutely continuous, implying it is differentiable almost everywhere with respect to $v_i^t$. Moreover, using (b) of Lemma 1, we can establish the following theorem.

\begin{Theorem}
\normalfont \textit{Suppose $\Gamma=(a^t,p^t)_{t \in \mathcal{T}}$ is incentive compatible and feasible. Then at any period $t \in \mathcal{T}$,}
\begin{itemize}
\item[(a)] \textit{$\forall (i,t)$, $A_{i}^{t}(v_{i}^{t}, q_{i}^{t} \vert q^t)$ is non-decreasing in $v_{i}^{t}$ for fixed $q_{i}^{t}$.}
\item[(b)] $\displaystyle U_{i}^{t}(v_{i}^{t}, q_{i}^{t} \vert q^t) = U_{i}^{t}( \underline{v}, q_{i}^{t}\vert q^t) + \int_{\underline{v}}^{v_{i}^{t}}{ A_{i}^{t}(\tau, q_{i}^{t}\vert q^t) d\tau }$.
\end{itemize}    
\end{Theorem}
\textit{Proof.} 

\noindent (a) From (b) of Lemma 1, for any $\varepsilon > 0$, we have
\begin{multline*}
    \frac{U_{i}^{t}  (v_{i}^{t}, q_{i}^{t} \vert q^t)  -   U_{i}^{t}(v_{i}^{t} - \varepsilon, q_{i}^{t} \vert q^t)  }{  \varepsilon}
     \leq A_{i}^{t}(v_{i}^{t},q_{i}^{t} \vert q^t) \\ \leq \frac{   U_{i}^{t}(v_{i}^{t}+ \varepsilon, q_{i}^{t} \vert q^t) - U_{i}^{t}(v_{i}^{t}, q_{i}^{t} \vert q^t)   }{  \varepsilon}
\end{multline*}

If $\varepsilon \rightarrow 0$, we get $\displaystyle \partial U_{i}^{t}(v_{i}^{t}, q_{i}^{t} \vert q^t)  / \partial v_{i}^{t} = A_{i}^{t}(v_{i}^{t},q_{i}^{t}\vert q^t)$ almost every everywhere and since $U_{i}^{t}(v_{i}^{t}, q_{i}^{t}\vert q^t)$ is convex in $v_{i}^{t}$, $A_{i}^{t}(v_{i}^{t},q_{i}^{t}\vert q^t)$ is non-decreasing in $v_{i}^{t}$.  

\noindent (b) Since $\displaystyle \partial U_{i}^{t}(v_{i}^{t}, q_{i}^{t} \vert q^t)  / \partial v_{i}^{t} = A_{i}^{t}(v_{i}^{t},q_{i}^{t}\vert q^t)$ almost everywhere, by the fundamental theorem of calculus, 
\begin{align*}
    \int_{\underline{v}}^{v_{i}^{t}}{A_{i}^{t}(\tau, q_{i}^{t} \vert q^t) d\tau } & = \int_{\underline{v}}^{v_{i}^{t}}{\dfrac{\partial U_{i}^{t}(\tau, q_{i}^{t}\vert q^t)  }{\partial v_{i}^{t}}  d\tau } \\ & =  U_{i}^{t}(v_{i}^{t},q_{i}^{t}\vert q^t) - U_{i}^{t}(\underline{v},q_{i}^{t}\vert q^t),
\end{align*}
which completes the proof. \hfill$\blacksquare$

Theorem 1 outlines the essential properties that any incentive compatible mechanism should satisfy. Particularly, using (b) of Theorem 1, the original revenue maximization problem is transformed into a problem analogous to Myerson's virtual value maximization problem. The virtual valuation of the two-dimensional bids is defined as follows

\begin{Definition} \normalfont \textbf{(Virtual Valuation)} \textit{For any realized buyer $(i,t)$ with $\theta_i^t = (v_{i}^{t},q_{i}^{t})$, the virtual valuation of bidder $(i,t)$ is defined as $\phi_{i}^{t}(\theta^t_i):=v_i^t - \frac{1 - F(v_i^t \vert q_i^t)}{f(v_i^t \vert q_i^t)}$.}
\end{Definition}

\begin{Theorem} \normalfont $(\clubsuit)$
\textit{For an incentive compatible, individually rational, and feasible mechanism $\Gamma = (a,p)$, the seller's problem is reduced to the following dynamic stochastic knapsack problem:} 
\begin{align}
\underset{a}{\max} & \quad \sum_{t=1}^{T} \delta^{t-1} \mathbb{E}_{n^t, \theta^{t}} \left[ \sum_{i = 1}^{n^t}{   \phi_i^t (\theta^t_i) a_{i}^{t}(\theta^{t}, q^t) }  \right] \\
\mbox{s.t.}    & \quad (5), \ (6) \nonumber
\end{align}
\end{Theorem}

Subsequently, the dynamic programming approach can be employed to solve the proposed stochastic program. For every possible state $(\theta^t, q^t)$, define $V^t(\theta^t, q^t)$ as the state value function, which represents the discounted sum of values that can be obtained from state $(\theta^t, q^t)$ until the end of the study period by following the optimal policy. Thus, the optimal solution must satisfy the following Bellman equation for every $t \in \mathcal{T}$ almost everywhere with respect to the underlying probability space. 
\begin{multline}
V^t(\theta^t, q^t) = \underset{0 \le x \le q^t}{\sup} \{ R^t (\theta^t, q^t, x) \\ + \delta \mathbb{E}_{n^{t+1}, \theta^{t+1}}\left[ V^{t+1}(\theta^{t+1}, q^t-x)\right] \},    
\end{multline}
with the boundary condition $V^{T+1}(\theta^{T+1}, q^{T+1}) = 0$ for every $\theta^{T+1} \in \Theta^{N}$ and $q^{T+1} \in [0, \bar{Q}]$. Also, $R^t(\theta^t, q^t, x)$ represents the maximum periodic revenue attainable at state $(\theta^t, q^t)$ when the seller opts to sell $x$ units, which has a form of
\begin{align*}
    R^t (\theta^t, q^t, x) = \underset{a^t}{\max} & \quad \sum_{i=1}^{n^t} \phi_i^t a_i^t(\theta^t, q^t) \\
    \mathrm{s.t.} & \quad 0 \le a^t_i(\theta^t , q^t) \le q_i^t, \quad \forall(i,t) \\
    & \quad \sum_{i=1}^{n^t}a_i^t(\theta^t, q^t) = x.
\end{align*}

Since the suggested problem is a linear knapsack problem, the optimal solution is to allocate in descending order of virtual valuation. Consequently, for a given state $(\theta^t, q^t)$, the buyers can be reordered based on their virtual valuations, denoted as $\phi^t_{[1]} \ge \cdots \ge \phi^t_{[n^t]}$. Let the order of buyer $(i,t)$ be represented as $[i]$. Then, we have the following theorem.
    \begin{Theorem}
     If the seller decides to sell $x$ units, the optimal allocation rule is
    \begin{equation}
        (a^t_i)^{\ast}(\theta^t, q^t) = 
        \begin{cases}
        q_i^t & \mbox{if } [i] \le i^{\ast}(\theta^t, x) \\
        x - \sum_{j=1}^{i^{\ast}(\theta^t, x)}q_{[j]}^{t} & \mbox{if } [i]  = i^{\ast}(\theta^t, x) + 1 \\
        0 & \mbox{otherwise}
        \end{cases}
    \end{equation}
    where $i^{\ast}(\theta^t, x)$ is the integer that satisfies $\sum_{j=1}^{i^{\ast}(\theta^t, x)}{q_{[j]}^t} \le x < \sum_{j=1}^{i^{\ast}(\theta^t, x)+1}{q_{[j]}^t}$.
    \end{Theorem}
\textit{Proof.} 
Without loss of generality, assume $\phi_{i}^{t} \ge 0$ for every bidder $(i,t)$. The objective function value of the suggested problem is $\sum_{j=1}^{i^{\ast}(\theta^t, x)}\phi_{[j]}^t q_{[j]}^t + \phi_{[i^{\ast}(\theta^t, x)+1]}(x - \sum_{j=1}^{i^{\ast}(\theta^t, x)}q_{[j]}^t)$. Then, consider the dual problem:
\begin{align*}
    \underset{\lambda \ge 0, \ \mu}0{\min} & \quad \sum_{i=1}^{n^t}q_{i}^{t}\lambda_i + x \mu \\
    \mbox{s.t.} & \quad \lambda_i^t + \mu \ge \phi_i^t, \quad \forall (i,t).
\end{align*}   
Since $\lambda_i^t = \max\{0, \phi_i^t - \mu\}$, the dual problem is reduced to
\begin{align*}
    \underset{\mu}{\min} \quad \sum_{i=1}^{n^t}q_{i}^{t}\max\{0, \phi_i^t - \mu\} + x \mu
\end{align*}
If we set $\mu^{\ast} = \phi_{[i^{\ast}(\theta^t, x)]}^t$, then the objective value of the dual problem is the same as that of the primal problem, which satisfies the strong duality.\hfill$\blacksquare$

As depicted in (b) of Theorem 1 and Theorem 3, the optimal allocation rule and payment rule are determined for a given $x$. Therefore, we are left with the problem of determining the appropriate amount to sell in each period, denoted as $x^{\ast}(\theta^t, q^t)$. In order to achieve this, the following is essential.

\begin{Lemma} \normalfont $(\clubsuit)$ \textit{For a given state $(\theta^t, q^t)$,}

\begin{itemize}
    \item [(a)] \textit{$R^t(\theta^t, q^t, x)$ is non-decreasing with respect to $x$.} 
    \item [(b)] \textit{$R^t(\theta^t, q^t, x)$ is concave with respect to $x$.}
    \item [(c)] \textit{$R^t(\theta^t, q^t, x)$ is concave with respect to $(q^t, x)$.}
\end{itemize}
\end{Lemma}

\begin{Lemma} \normalfont $(\clubsuit)$ \textit{For a given state $(\theta^t, q^t)$,}
$\,$
\begin{itemize}
    \item [(a)] \textit{$V^t(\theta^t, q^t)$ is non-decreasing in $q^t$.}
    \item [(b)] \textit{$V^t(\theta^t, q^t)$ is concave with respect to $q^t$}
    \item [(c)] \textit{$\mathbb{E}_{n^t, \theta^t}[V^t(\theta^t, q^t)]$ is non-decreasing and concave with respect to $q^t$}
\end{itemize}
\end{Lemma}

From Lemma 2 and Lemma 3, we can notice that the functions $R^t(\theta^t, q^t, x)$ and $\mathbb{E}_{n^{t+1}, \theta^{t+1}}[V^{t+1}(\theta^{t+1}, q^t - x)]$ are monotone with respect to $x$. As a result, they are differentiable almost everywhere. Therefore, the differentiation of the state value function is well-defined.

\begin{Definition} \normalfont \textbf{(Marginal Value)} \textit{For any state $(\theta^t, q^t)$, the marginal value, denoted as $MV^t(\theta^t, q^t, x)$, is defined as}
\begin{multline}
     MV^t(\theta^t, q^t, x) = \frac{\partial}{\partial x} \{R^t(\theta^t, q^t, x) + \\ \delta \mathbb{E}_{n^{t+1}, \theta^{t+1}}[V^{t+1}(\theta^{t+1}, q^t - x)] \}.
\end{multline}
\end{Definition}
Note that $MV^t(\theta^t, q^t, x)$ is non-increasing with respect to $x$ since both $R^t(\theta^t, q^t, x)$ and $\mathbb{E}_{n^{t+1}, \theta^{t+1}}[V^{t+1}(\theta^{t+1}, q^t - x)]$ are concave. Also, they are continuous almost everywhere, which makes them Riemann integrable. Given that $R^t(\theta^t, q^t, 0) = 0$, the Bellman equation can be rewritten as
\begin{multline}
V^{t}(\theta^t, q^t) = \underset{0 \le x \le q}{\sup} \left\{ \int_{0}^{x} MV^t(\theta^t, q^t, \tau) d \tau \right\} + \\ \delta \mathbb{E}_{n^{t+1}, \theta^{t+1}}[V^{t+1}(\theta^{t+1}, q^{t})], 
\end{multline}
which implies $x^{\ast}(\theta^t, q^t) \in \underset{0 \le x \le q^t}{\argsup} \int_{0}^{x} MV^t(\theta^t, q^t, \tau) d \tau$. 

Then, define the set $X^{-}(\theta^t, q^t)$ that includes $x$ with negative marginal values in a given state $(\theta^t, q^t)$, i.e., 
\begin{equation}
    X^{-}(\theta^t, q^t) = \{x \in [0, q^t] \vert MV^t(\theta^t, q^t, x) < 0 \}.
\end{equation}
If the set $X^{-}(\theta^t, q^t)$ is non-empty, the infimum is well-defined because 0 is a lower bound. The following theorem proposes the optimal $x^{\ast}(\theta^t, q^t)$.

\begin{Theorem}
    For a given state $(\theta^t, q^t)$, $x^{\ast}(\theta^t, q^t)$ is
\begin{equation}
    x^{\ast}(\theta^t, q^t) = 
    \begin{cases}
        q^t & \mbox{if }  X^{-}(\theta^t, q^t) = \varnothing  \\
        \inf X^{-}(\theta^t, q^t) & \mbox{otherwise}.
    \end{cases}
\end{equation}
\end{Theorem}
\textit{Proof.} 
Let a state $(\theta^t, q^t)$ be given. First, consider the case where $X^{-}(\theta^t, q^t) = \varnothing$. Then, for all $x \in [0, q^t]$, $MV^t(\theta^t, q^t, x) \ge 0$. So, for all $x \in [0, q^t]$, 
\begin{align*}
\int_{0}^{x}MV^{t}(\theta^t, q^t, \tau)d\tau \le  \int_{0}^{q^t}MV^{t}(\theta^t, q^t, \tau)d\tau. 
\end{align*}
Therefore, $x^{\ast}(\theta^t, q^t) = q^t$ in this case. Then, consider the case where  $X^{-}(\theta^t, q^t) \neq \varnothing$. Denote $x^{\circ} = \inf X^{-}(\theta^t, q^t)$. If $x^{\circ} = 0$, it is obvious that $x^{\circ}$ is optimal because $MV^t(\theta^t, q^t, x) < 0$ for every $x \in (0, q^t]$. If $x^{\circ} > 0$, we have $MV^t(\theta^t, q^t, x) \ge 0$ for $0 \le x < x^{\circ}$ and $MV^t(\theta^t, q^t, x) < 0$ for $x^{\circ} < x \le q^t$ since $MV^t(\theta^t, q^t, x)$ is non-increasing with respect to $x$. Assume that $x^{\circ} \notin \argsup \int_{0}^{x} MV^t(\theta^t, q^t, \tau) d \tau$. Then, there exists $x'$ that attains the supremum and $x' \neq x^{\circ}$. If $0 \le x' < x^{\circ}$, 
\begin{align*}
     \int_{0}^{x'} MV^t(\theta^t, q^t, \tau) d \tau 
      \le \int_{0}^{x^{\circ}} MV^t(\theta^t, q^t, \tau) d \tau,
\end{align*}
which is a contradiction that $x^{\circ}$ does not attain the supremum. Similarly, if $x^{\circ} < x' \le q^t$,
\begin{align*}
     \int_{0}^{x'} MV^t(\theta^t, q^t, \tau) d \tau  < \int_{0}^{x^{\circ}} MV^t(\theta^t, q^t, \tau) d \tau,
\end{align*}
which is a contradiction that $x'$ attains the supremum. \hfill $\blacksquare$

Theorem 4 provides us with the optimal solution to the Bellman equation outlined in (8). However, it's crucial to remember that we have not taken into account the non-decreasing property of an incentive compatible mechanism, as presented in (a) of Lemma 1. To ensure the property of incentive compatibility, we define the following assumption.
\begin{Assumption} \normalfont \textbf{(Regularity condition)} \textit{The virtual valuation $\phi_i^t(\theta^t_i) = v_i^t - \frac{1 - F(v_i^t \vert q_i^t)}{f(v_i^t \vert q_i^t)}$ is non-decreasing with respect to $v_i^t$ and $q_i^t$.} 
\end{Assumption}

Under the regularity condition, buyers with a high value and a greater demanded quantity will have a higher priority if the seller determines to allocate the item as suggested in (9). Therefore, we can obtain the following theorem. 
\begin{Theorem} $(\clubsuit)$ Under the regularity condition, \normalfont
\begin{itemize}
    \item [(a)] \textit{$(a_i^t)^{\ast}(\theta^t_i, \theta^t_{-i}, q^t)$ and $(A_i^t)^{\ast}(\theta^t_i \vert q^t)$ are non-decreasing with respect to $v_i^t$}. 
    \item [(b)]  \textit{$(a_i^t)^{\ast}(\theta^t_i, \theta^t_{-i}, q^t)$ and $(A_i^t)^{\ast}(\theta^t_i \vert q^t)$ are non-decreasing with respect to $q_i^t$.} 
\end{itemize}
\end{Theorem}

Nevertheless, despite deriving the optimal allocation and payment rules based on the necessity condition of incentive compatibility and the regularity condition, it remains to be proven that the proposed mechanism is genuinely incentive compatible. Specifically, it is found that if the mechanism solely consists of the derived allocation and payment rules, it is found that buyers can overbid on their demanded quantities to increase their purchase probability. To prevent this, we develop a penalty scheme that can punish buyers when such quantity overbidding occurs.

\begin{Assumption}
After the allocation, the seller can observe whether the allocated quantity $(a_i^t)^{\ast}(\hat{\theta}^t_i, \theta^t_{-i}, q^t)$ is greater than the true demanded quantity $q_i^t$, for every buyer at no cost.
\end{Assumption}
This assumption indicates that although the seller might not know the buyers' desired quantities initially, they can ascertain them after the bidding and allocation are finished. This assumption is reasonable in the field of rental or service businesses, which is the main motivation of this study, where the system can observe whether allocated resources are actually being utilized after the allocation has taken place. We define the penalty scheme as follows.

\begin{Definition}\normalfont \textbf{(Penalty Scheme)}
\textit{If a bidder $(i,t)$ who reports $\hat{\theta}^t_i = (v_i^t, \hat{q}_i^t)$ overbids the quantity, i.e., $\hat{q}_i^t > q_i^t$, and subsequently receives an allocation greater than their true quantity, i.e., $a_i^t(\hat{\theta}^t_i, \theta^t_{-i}, q^t) > q_i^t$, the bidder must pay a certain amount of penalty defined as}
\begin{align}
    \rho_i^t(\hat{\theta}^t_i, \theta^t_{-i}, q^t) = 
     \frac{\bar{v} \hat{q}_i^t}{\mathcal{P}_{\theta^{t}_{-i}}((a_i^t)^{\ast, -1}(\{\hat{q}_i^t\}))}, 
\end{align}
\textit{where $\mathcal{P}_{\theta^{t}_{-i}}((a_i^t)^{\ast, -1}(\{\hat{q}_i^t\}))$ is the probability that the allocated quantity is equal to $\hat{q}_i^t$, i.e., $(a_i^t)^{\ast}(\hat{\theta}^t_i, \theta^t_{-i}, q^t) = \hat{q}_i^t $.}
\end{Definition}

With the defined penalty scheme, we prove that the proposed mechanism is incentive compatible, individually rational, and achieves ex-ante revenue maximization. 

\begin{Theorem} \normalfont $(\clubsuit)$
\textit{Under the regularity condition, the following mechanism $\Gamma^{\ast} = (a^{\ast}, p^{\ast})$ with the penalty scheme $\rho$ satisfies the incentive compatibility, individual rationality, and the optimality:
\begin{equation*}
    (a_i^t)^{\ast}(\theta^t, q^t) = 
    \begin{cases}
        q_i^t & \mbox{if } [i] \le i^{\ast} \\
        x - \sum_{j=1}^{i^{\ast}}q^t_{[j]} & \mbox{if } [i] = i^{\ast}+1 \\
        0 & \mbox{otherwise}
    \end{cases}, 
\end{equation*}
\small
\begin{equation*}
    (p_i^t)^{\ast}(\theta^t, q^t) = v_i^t (a_i^t)^{\ast}(\theta^t, q^t) - \int_{\underline{v}}^{v_i^t} (a_i^t)^{\ast}((\tau, q_i^t), \theta^t_{-i}, q^t) d\tau    
\end{equation*}
\normalsize
where $i^{\ast} = i^{\ast}(\theta^t, x^{\ast}(\theta^t, q^t))$ is the integer that satisfies $\sum_{j=1}^{i^{\ast}(\theta^t, x^{\ast}(\theta^t, q^t))}q^t_{[j]} \le x^{\ast}(\theta^t, q^t) < \sum_{j=1}^{i^{\ast}(\theta^t, x^{\ast}(\theta^t, q^t)) + 1}q^t_{[j]}$ and 
\begin{equation*}
    x^{\ast}(\theta^t, q^t) = 
    \begin{cases}
        q^t & \mbox{if } X^{-}(\theta^t, q^t) = \varnothing \\
        \inf X^{-}(\theta^t, q^t) & \mbox{otherwise}.
    \end{cases}
\end{equation*}}
\end{Theorem}

\section{Approximation Algorithms}
The optimal mechanism that we have derived comprises the allocation rule, payment rule, and $x^{\ast}(\theta^t, q^t)$, which serves to determine the allocation amount for each period. However, the practical implementation of this mechanism necessitates a computational approach. The allocation rule and payment rule entail a computational complexity of $O(N \log N)$ due to the need to sort bidders by their virtual valuations at each time period. Further details of these algorithms are provided in the supplementary material. The complexity arises in computing $x^{\ast}(\theta^t, q^t)$ due to the requirement to calculate the expected value of the state value function in a continuous space. To tackle this challenge, we propose two algorithms designed to approximate $x^{\ast}(\theta^t, q^t)$. 

\subsection{Approximation Methods} 
We then present two algorithms to approximate $x^{\ast}(\theta^t, q^t)$. To do this, we first consider approximating the expectations of the state value function using the MC simulation-based regression. Assume that there are some basis functions $f_1, \cdots, f_n$. Then, the state value function is approximated as a linear combination of the basis functions: $V^t(\theta^t, q^t) = V^t(q^t) \approx \sum_{j=0}^{n} c^t_{j} f_j(q^t)$. The coefficients $c^t_j$ are determined through regression at some fixed states $s_{1}, \cdots, s_{m}$. In this study, we use the Chebyshev polynomials and nodes as basis functions and states:
\begin{equation}
    f_0(s) = 1, \ f_1(s) = s, \ f_{n+1}(s) = 2sf_n(s) - f_{n-1}(s),
\end{equation}
\begin{equation}
    s_{k} = \frac{\bar{Q}}{2}\left\{ 1 + \cos \left( \frac{2k-1}{m} \pi \right) \right\}, \ \mbox{for } k = 1,\cdots,m.
\end{equation}
Note that the Chebyshev polynomials have the characteristics that they are orthogonal functions, i.e., $\int_{-1}^{1}f_{i}(s)f_{j}(s) = 0$ for any $i \neq j$. Also, the original Chebyshev nodes are $\cos \left( \frac{2k-1}{m} \pi \right)$ which ranges from -1 to 1, so they are resized from 0 to $\bar{Q}$. Then, the original Bellman equation is reduced to the following equation: 
\begin{align}
    \sum_{j=0}^{n}c_j^t f_j(s_k)  \approx \underset{x \le s_k}{\max} \ R^t(\theta^t, s_k, x) + \delta \sum_{j=1}^{n}c_j^{t+1} f_j(s_k - x). 
\end{align}
After obtaining the coefficients, the contingent decisions can be made by the approximated state value functions. The simulation algorithm is presented in Algorithm \ref{alg:MC}. 

\begin{algorithm}[t]
\normalsize
\caption{Monte Carlo Simulation}
\label{alg:MC}
\textbf{Input}: $n$ (\# of basis); $m$ (\# of nodes); $N^{e}$ (\# of episodes)

\begin{algorithmic}[1] 

\STATE{Initialize $V^t(s_k) = 0$, $t = 1,\cdots, T$, $k = 1, \cdots, m$}
\STATE{Initialize coefficient matrix $C \in \mathbb{R}^{T \times (n+1)} = \mathbf{0}$}
\STATE{$i = 1$}
\WHILE{$i \le N^{e}$}
\FOR{$t = T, \cdots, 1$}
\STATE{Generate buyers' profiles $\theta^t$}
\FOR{$s = s_1, \cdots, s_m$}
\STATE{$x^{\ast} = \argmax_{x \le s} \{ R^t(\theta^t, s, x) + \delta V^{t+1}(s-x) \}$}
\STATE{$V^{new} = R^t(\theta^t, s, x^{\ast}) + \delta V^{t+1}(s - x^{\ast})$}
\STATE{$V^t(s) \leftarrow \frac{i-1}{i}V^t(s)  + \frac{1}{i} V^{new}$}
\ENDFOR
\STATE{$i \leftarrow i+1$}
\ENDFOR
\ENDWHILE
\FOR{$t = 1, \cdots, T$}
\STATE{Fit $\begin{bmatrix} f_0(s_1) & \cdots & f_n(s_1) \\ \vdots & \ddots & \vdots \\ f_0(s_m) & \cdots & f_n(s_m)  \end{bmatrix}  \begin{bmatrix} c^t_1 \\ \vdots \\ c^t_n \end{bmatrix}\approx \begin{bmatrix} V^t(s_1) \\ \vdots \\ V^t(s_m) \end{bmatrix}$}
\ENDFOR
\STATE \textbf{return} $C$ 
\end{algorithmic}
\end{algorithm}

Meanwhile, an alternative approach is to approximate the optimal policy through policy learning. Specifically, reinforcement learning serves as a technique to learn the optimal policy within a dynamic environment. In our case, as the model encompasses a continuous space, we propose the DDPG method. This actor-critic method is well-suited for handling continuous action spaces. A concise overview of the DDPG algorithm pertaining to dynamic allocation is provided in Algorithm \ref{alg:DDPG}. For an in-depth exploration of this algorithm, please refer to  \citet{lillicrap2015continuous}. 

\begin{algorithm}[t]
\normalsize
\caption{Deep Deterministic Policy Gradient method}
\label{alg:DDPG}

\textbf{Input}: $\tau$ (learning rate), $N^e$, $r$ (\# of random episodes)

\begin{algorithmic}[1] 

\STATE{Initialize critic $Q(s,a \vert \theta^Q)$ and actor $\mu(s \vert \theta^{\mu})$}
\STATE{Initialize target critic ($Q'$) and actor ($\mu'$) with $\theta^Q$ and $\theta^{\mu}$}
\STATE{Initialize Replay buffer $R$}
\FOR{$i = 1,\cdots, N^e$}
\IF{$i \le r$}
\FOR{$t = 1, \cdots, T$}
\STATE{At state $s_t = q^t$, generate buyers' profile $\theta^t$}
\STATE{Agent chooses random action $a_t = x$}
\STATE{Compute reward $r_t = R^t(\theta^t, q^t, x)$}
\STATE{Compute the next state $s_{t+1} = s_t - x$}
\STATE{Store $(s_t, a_t, r_t, s_{t+1})$ in replay buffer $R$}
\ENDFOR
\ELSE
\STATE{Minibatch $(s_t, a_t, r_t, s_{t+1})$ from replay buffer}
\STATE{Train critic and actor network, $\theta^Q$ and $\theta^{\mu}$}
\STATE{Update target networks: \\ $\theta^{Q'} \leftarrow \tau \theta^{Q} + (1 - \tau)\theta^{Q'}$, $\theta^{\mu'} \leftarrow \tau \theta^{\mu} + (1 - \tau)\theta^{\mu'}$}
\FOR{$t=1,\cdots, T$}
\STATE{Choose action $a_t = \mu(s_t \vert \theta^{\mu}) + \epsilon_t$}
\STATE{Store $(s_t, a_t, r_t, s_{t+1})$ in replay buffer $R$}
\ENDFOR
\ENDIF
\ENDFOR
\STATE \textbf{return} $\mu$, $\mu'$, $Q$, $Q'$ 
\end{algorithmic}
\end{algorithm}

\subsection{Numerical Experiment} 
To compare the effectiveness of the two proposed approximation algorithms, we conduct numerical experiments under various scenarios, especially the length of the study period. We consider three different market environment where $(T. \bar{Q}) = (10, 10)$, $(30, 30)$, and $(100, 100)$. The distribution of the buyers and the seller's plans are determined as shown in Table \ref{table:param}. 
\begin{table}[b]
\centering
\resizebox{\columnwidth}{!}{%
\begin{tabular}{ll}
\hline
Parameters                     & Values                                                                                  \\ \hline
$g(n)$                  & Poisson(10) $(g(n) = \frac{10^{n}e^{-10}}{n!})$  \\
$f(q)$            & Uniform(0, 2) $(f(q)=\frac{1}{2}$ for $0 \le q \le 2)$                                                \\
$f(v \vert q)$ & Exponential($q$) $(f(v \vert q) = qe^{-qv})$                                                               \\
$(T, \bar{Q})$     & (10, 10), (30, 30), (100, 100)                                                          \\
$\delta$                 & 0.99                                                            \\ \hline
\end{tabular}%
}
\caption{Experimental parameters}
\label{table:param}
\end{table}
For the simulation, we employ 5 basis functions and vary the number of nodes ($m$) within the range of 5 to 50. In the context of DDPG, we set the learning rates for the actor, critic, and soft update to 0.0001, 0.001, and 0.0001 respectively. A minibatch size of 64 is utilized, and the neural network structure encompasses 3 layers, each with 64 nodes. Also, the initial random choices are set to be 10\% of the total training episodes. All the methods are trained by 10,000 episodes and the performances are compared by averaging 20 test episodes. The simulations were performed on a computer with an Intel Core i7-6700 CPU, 16GB RAM, and an NVIDIA GeForce GTX 1060 GPU, using Python numpy and Pytorch packages, with a fixed seed number of 1.

The results, including the average discounted rewards from test episodes and their corresponding training times, are presented in Table \ref{table:results}. The full information case that maximizes the virtual value is presented together to suggest theoretical upper bounds. The MC method outperforms the DDPG approach in terms of average test rewards, even when the number of nodes in the MC method is kept small. However, While increasing the number of nodes in the MC method enhances performance, it considerably slows down training speed. Conversely, DDPG exhibits notably faster computation speed in comparison to the MC method.

\begin{table}[t]
\centering
\resizebox{\columnwidth}{!}{%
\begin{tabular}{llll}
\hline
Methods & \multicolumn{3}{l}{Average test rewards (training time: min)} \\\cline{2-4} 
        & $(10, 10)$ & $(30, 30)$ & $(100, 100)$   \\ \hline
MC ($m=5$)          & 21.77 (0.9)    & 56.71 (2.7)       & 115.88 (9.1) \\
MC ($m=10$)         & 21.80 (3.0)    & 62.84 (9.3)       & 132.62 (34.5) \\
MC ($m=20$)         & 21.88 (11.0)   & 63.48 (34.5)      & 146.75 (119.4) \\
MC ($m=50$)         & 21.88 (65.4)   & 63.74 (208.9)     & 163.40 (714.4) \\
DDPG                & 13.40 (1.0)    & 32.16 (1.9)       & 91.24 (5.1) \\
Full information    & 22.78          & 65.23    & 170.62      \\\hline
\end{tabular}%
}
\caption{Average test rewards and training time of the approximation methods}
\label{table:results}
\end{table}

\begin{figure}[b]
\centering
\includegraphics[width=0.9\columnwidth]{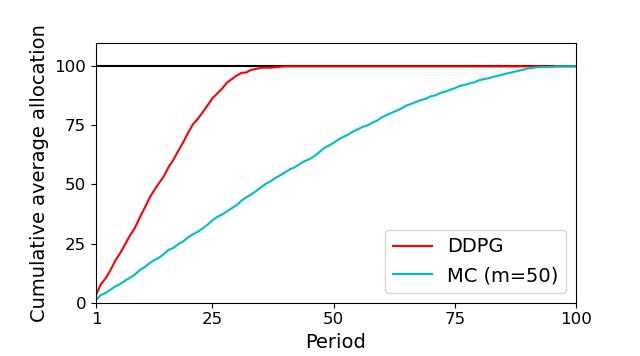} 
\caption{Cumulative allocation of $(T, \bar{Q}) = (100, 100)$}
\label{fig1}
\end{figure}

To investigate the reasons behind DDPG's lower performance compared to the MC method, we analyze the cumulative average allocation depicted in Figure \ref{fig1}. Notably, the MC method effectively adjusts the distribution of item quantities across different time periods, while DDPG exhibits a tendency to allocate items to buyers arriving early in the study period. This phenomenon can be attributed to two main factors. Firstly, due to the inherent characteristics of DDPG's model, there might be an overestimation of the value function, resulting in an early allocation of resources. Secondly, this premature allocation results from a lack of sufficient state generation for learning during the later stages. 

\section{Conclusion}
We designed the optimal dynamic mechanism and suggested corresponding approximation algorithms under a more generalized dynamic stochastic knapsack environment. We figured out that the penalty scheme is necessary to preserve incentive compatibility under the regularity condition on a two-dimensional type. Meanwhile, the presented approximation algorithms showed that the DDPG-based method has a lower performance than the regression method. To address these issues, it is evident that improvements in exploration during the policy learning process of DDPG are necessary to train the model using allocation data from the later stages.

\section{Acknowledgments}
This work was supported by Basic Science Research Program through the National Research Foundation of Korea (NRF) funded by the Ministry of Education (2021R1\(|\)1A4A01059254). 

\bibliography{aaai24}

\end{document}

\title{Appendix}
\author{}
\date{}
\maketitle 
\subsection*{Proof of Lemma 1}
\begin{proof}[\unskip\nopunct]

\noindent (a) For any feasible mechanism $\Gamma = (a,p)$, if the bidder whose true type is $\theta_i^t = (v_i^t, q_i^t)$ reports its type as $\hat{\theta}_i^t = (\hat{v}_i^t, \hat{q}_i^t)$, the seller will allocate the goods less than the reported demand, i.e., $a_i^t((\hat{v}_i^t, \hat{q}_i^t), \theta^t_{-i}, q^t) \le \hat{q}_i^t$. 

Then, if the bidder reports its demand truthfully, i.e., $\hat{q}_i^t = q_i^t$, we can rewrite the interim utility of the bidder as  
\begin{align*}
U_{i}^{t}((\hat{v}_i^t, q_i^t) & \vert (v_i^t, q_i^t), q^t)   \\ & = \mathbb{E}_{\theta_{-i}^t}[v_i^t \min\{q_i^t, a_i^t((\hat{v}_i^t, q_i^t), \theta^t_{-i}, q^t)\} - p_i^t((\hat{v}_i^t, q_i^t), \theta^t_{-i}, q^t)]  \\
& = 
\mathbb{E}_{\theta_{-i}^t}[v_i^t a_i^t((\hat{v}_i^t, q_i^t), \theta^t_{-i}, q^t) - p_i^t((\hat{v}_i^t, q_i^t), \theta^t_{-i}, q^t)]   \\
& = v_i^t A_i^t(\hat{v}_i^t, q_i^t \vert q^t) - P_i^t(\hat{v}_i^t, q_i^t \vert q^t).
\end{align*}
Meanwhile, since $\Gamma=(a,p)$ is incentive compatible, truth-telling maximizes the interim utility of the bidder and we have 
\begin{align}
    U_i^t(v_i^t, q_i^t \vert q^t) & = v_i^t A_i^t(v_i^t, q_i^t \vert q^t) - P_i^t(v_i^t, q_i^t \vert q^t) \nonumber\\
    & = \underset{\hat{v}_i^t \in [\underline{v}, \bar{v}]}{\sup} v_i^t A_i^t(\hat{v}_i^t, q_i^t \vert q^t) - P_i^t(\hat{v}_i^t, q_i^t \vert q^t). 
\end{align}

Then, let $v_{i}^{t} = \alpha v_{1} + (1 - \alpha) v_{2}$ where  $v_{1}$, $v_{2}  \in [\underline{v}, \bar{v}]$ and $\alpha \in [0,1]$ are given. Then the following holds. 
\begin{align*}
    U_{i}^{t}(v_{i}^{t}, q_{i}^{t} \vert q^t) &= v_{i}^{t}A_{i}^{t}(v_{i}^{t}, q_{i}^{t} \vert q^t) - P_{i}^{t}(v_{i}^{t}, q_{i}^{t} \vert q^t) \\
    &= \{\alpha v_{1} + (1 - \alpha) v_{2}\}A_{i}^{t}(v_{i}^{t}, q_{i}^{t} \vert q^t) - P_{i}^{t}(v_{i}^{t}, q_{i}^{t}\vert q^t) \\
    &= \alpha\{v_{1}A_{i}^{t}(v_{i}^{t}, q_{i}^{t}\vert q^t) - P_{i}^{t}(v_{i}^{t}, q_{i}^{t}\vert q^t)\} \\
    & \qquad +(1 - \alpha)\{v_{2}A_{i}^{t}(v_{i}^{t}, q_{i}^{t}\vert q^t) - P_{i}^{t}(v_{i}^{t}, q_{i}^{t}\vert q^t)\} \\
    & \leq \alpha U_{i}^{t}(v_{1}, q_{i}^{t}\vert q^t) +  (1 - \alpha) U_{i}^{t}(v_{2}, q_{i}^{t}\vert q^t). 
\end{align*}
The last inequality holds due to (2). 

\,\

\noindent (b) We only prove the left part of the inequality. The right part of the inequality is omitted and can be proved similarly. For any $(v_{i}^{t}, q_{i}^{t}) \in \Theta$ and $\varepsilon>0$, we have
\begin{align*}
U_{i}^{t}(v_{i}^{t} - \varepsilon, q_{i}^{t} \vert q^t) & \geq U_{i}^{t}( (v_{i}^{t}, q_{i}^{t} ) \vert (v_{i}^{t}- \varepsilon, q_{i}^{t}  ) , q^t ) &\\
& = (v_{i}^{t}- \varepsilon ) A_{i}^{t}(v_{i}^{t},q_{i}^{t} \vert q^t)- P_{i}^{t}(v_{i}^{t},q_{i}^{t}\vert q^t) &&\\
& = v_{i}^{t}A_{i}^{t}(v_{i}^{t},q_{i}^{t}\vert q^t) - P_{i}^{t}(v_{i}^{t},q_{i}^{t}\vert q^t)- \varepsilon A_{i}^{t}(v_{i}^{t},q_{i}^{t}\vert q^t) &&\\
& = U_{i}^{t}(v_{i}^{t}, q_{i}^{t} \vert q^t) - \varepsilon A_{i}^{t}(v_{i}^{t},q_{i}^{t} \vert q^t),
\end{align*}
which is equal to $U_{i}^{t}(v_{i}^{t}, q_{i}^{t}\vert q^t) - U_{i}^{t}(v_{i}^{t} - \varepsilon, q_{i}^{t}\vert q^t) \leq \varepsilon A_{i}^{t}(v_{i}^{t},q_{i}^{t} \vert q^t)$. 

\end{proof}

\subsection*{Proof of Theorem 2}
\begin{proof}[\unskip\nopunct]

\noindent From (b) of Theorem 1, under an incentive compatible and feasible mechanism, the expected payment can be written as
\begin{equation*}
P_{i}^{t}(\theta^{t}_{i} \vert q^t) = v_i^tA_i^t(\theta^{t}_{i}\vert q^t) - \int_{\underline{v}}^{v_{i}^{t}}A_{i}^{t}(\tau, q_{i}^{t}\vert q^t)d\tau - U_{i}^{t}(\underline{v}, q_{i}^{t}\vert q^t). 
\end{equation*}
If we substitute it into the ex-ante payment of bidder $i$, $\mathbb{E}_{\theta^t}\left[ p_{i}^{t}(\theta^t \vert q^t) \right]$ equals
\begin{align*}
  &\mathbb{E}_{\theta^t_i}\left[ P_{i}^{t}(\theta^t_i \vert q^t) \right] \\
    &= \int_{\underline{q}}^{\bar{q}}\int_{\underline{v}}^{\bar{v}} \left(  v_i^t A_i^t(v_i^t,q_i^t \vert q^t) - \int_{\underline{v}}^{v_i^t}A_{i}^{t}(\tau, q_i^t \vert q^t)d\tau - U_{i}^{t}(\underline{v}, q_i^t \vert q^t)\right)f(v_i^t,q_i^t )dv_i^tdq_i^t \\
    &= \int_{\underline{q}}^{\bar{q}}\int_{\underline{v}}^{\bar{v}}\left\{ \left( v_{i}^{t} - \dfrac{1-F_{i}( v_{i}^{t} \rvert q_{i}^{t}) }{f_{i}(v_{i}^{t} \rvert q_{i}^{t})  } \right) A_i^t(v_i^t,q_i^t \vert q^t) -  U_{i}^{t}(\underline{v}, q_i^t \vert q^t) \right\} f(v_i^t,q_i^t)dv_i^tdq_i^t \\
    &= \mathbb{E}_{\theta^t}\left[\left( v_{i}^{t} - \dfrac{1-F_{i}( v_{i}^{t} \rvert q_{i}^{t}) }{f_{i}(v_{i}^{t} \rvert q_{i}^{t})  } \right)a_i^t(\theta^t \vert q^t) - U_{i}^{t}(\underline{v}, q_i^t \vert q^t)  \right].
\end{align*}
We can show the third equality by changing the order of integration, which can be computed as
\begin{align*}
\int_{\underline{q}}^{\bar{q}}\int_{\underline{v}}^{\bar{v}} \int_{\underline{v}}^{v_i^t} & A_{i}^{t}(\tau, q_i^t \vert q^t)d\tau f(v_i^t,q_i^t)dv_i^tdq_i^t \\ 
&= \int_{\underline{q}}^{\bar{q}}\int_{\underline{v}}^{\bar{v}} \int_{\tau}^{\bar{v}}f(v_i^t \vert q_i^t)dv_i^t A_{i}^{t}(\tau, q_i^t \vert q^t)f(q_i^t)d\tau dq_i^t \\
&= \int_{\underline{q}}^{\bar{q}}\int_{\underline{v}}^{\bar{v}} \left(1 - F(v_{i}^{t} \vert q_{i}^{t}) \right) A_{i}^{t}(v_i^t, q_i^t \vert q^t) f(q_i^t) dv_i^t dq_i^t \\
&= \int_{\underline{q}}^{\bar{q}}\int_{\underline{v}}^{\bar{v}} \left( \frac{1 - F(v_{i}^{t} \vert q_{i}^{t})}{f(v_{i}^{t} \vert q_{i}^{t})} \right) A_{i}^{t}(v_i^t, q_i^t \vert q^t) f(v_i^t, q_i^t) dv_i^t dq_i^t.
\end{align*}
Therefore, we can transform the objective function as follows. 
\begin{align*}
\sum_{t=1}^{T}\delta^{t-1} & \mathbb{E}_{n^t, \theta^t}\left[\sum_{i=1}^{n^t} p_{i}^{t}(\theta^t \vert q^t)\right] \\ &= \mathbb{E}_{n^t, \theta^{t}} \left[ \sum_{t=1}^{T}\sum_{i = 1}^{n^t}{  \delta^{t-1} \left( v_{i}^{t} - \dfrac{1-F_{i}( v_{i}^{t} \rvert q_{i}^{t}) }{f_{i}(v_{i}^{t} \rvert q_{i}^{t})  } \right) a_{i}^{t}(\theta^{t} \vert q^t) } - U_{i}^{t}(\underline{v}, q_i^t \vert q^t)   \right].
\end{align*}

Furthermore, as a profit maximizer, the seller can set $U_{i}^{t}(\underline{v}, q_i^t \vert q^t)  = 0$ for any $q_i^t \in [\underline{q}, \bar{q}]$ by adjusting a proper payment rule. 

\end{proof}

\subsection*{Proof of Lemma 2}
\begin{proof}[\unskip\nopunct]

\noindent (a) Recall that $R(\theta^t, q^t, x) = \sum_{j=1}^{i^{\ast}(\theta^t, x)}\phi_{[j]}^t q_{[j]}^t + \phi_{[i^{\ast}(\theta^t, x)+1]}(x - \sum_{j=1}^{i^{\ast}(\theta^t, x)} q_{[j]}^t )$. Let $x < y$, then it is clear that $i^{\ast}(\theta^t, x) \le i^{\ast}(\theta^t, y)$. 

If $i^{\ast}(\theta^t, x) = i^{\ast}(\theta^t, y)$, 
\begin{align*}
    R^t(\theta^t, q^t, x) & = \sum_{j=1}^{i^{\ast}(\theta^t, x)}\phi_{[j]}^t q_{[j]}^t + \phi_{[i^{\ast}(\theta^t, x)+1]}(x - \sum_{j=1}^{i^{\ast}(\theta^t, x)} q_{[j]}^t ) \\
    & < \sum_{j=1}^{i^{\ast}(\theta^t, x)}\phi_{[j]}^t q_{[j]}^t + \phi_{[i^{\ast}(\theta^t, x)+1]}(y - \sum_{j=1}^{i^{\ast}(\theta^t, x)} q_{[j]}^t ) \\
    & = R^t(\theta^t, q^t, y).
\end{align*}

If $i^{\ast}(\theta^t, x) < i^{\ast}(\theta^t, y)$, it follows that $i^{\ast}(\theta^t, x) + 1 \le i^{\ast}(\theta^t, y)$, and thus 
\begin{align*}
    R^t(\theta^t, q^t, x) & = \sum_{j=1}^{i^{\ast}(\theta^t, x)}\phi_{[j]}^t q_{[j]}^t + \phi_{[i^{\ast}(\theta^t, x)+1]}(x - \sum_{j=1}^{i^{\ast}(\theta^t, x)}q_{[j]}^t ) \\
    & < \sum_{j=1}^{i^{\ast}(\theta^t, x)+1}\phi_{[j]}^t q_{[j]}^t  \\
    & \le R^t(\theta^t, q^t, y).
\end{align*}

\noindent (b) Let $x < y$ and $\alpha \in (0,1)$. Denote $z = \alpha x + (1 - \alpha) y$. Then, since $x < z < y$, it follows that $i^{\ast}(\theta^t, x) \le i^{\ast}(\theta^t, z) \le i^{\ast}(\theta^t, y)$. Then, 
\begin{align*}
R^t(\theta^t, q^t, z)  = & \sum_{j=1}^{i^{\ast}(\theta^t, z)}{\phi_{[j]}^{t}q_{[j]}^t} + \phi_{[i^{\ast}(\theta^t, z) +1]}^t\left\{\alpha x + (1-\alpha) y - \sum_{j=1}^{i^{\ast}(\theta^t, z)}{q_{[j]}^t} \right\} \\
 = & \alpha \left\{ \sum_{j=1}^{i^{\ast}(\theta^t, z)}{\phi_{[j]}^{t}q_{[j]}^t} + \phi_{[i^{\ast}(\theta^t, z) +1]}^t \left(x - \sum_{j=1}^{i^{\ast}(\theta^t, z)}{q_{[j]}^t} \right)\right\} \\
 & + (1 - \alpha) \left\{ \sum_{j=1}^{i^{\ast}(\theta^t, z)}{\phi_{[j]}^{t}q_{[j]}^t} + \phi_{[i^{\ast}(\theta^t, z) +1]}^t \left(y - \sum_{j=1}^{i^{\ast}(\theta^t, z)}{q_{[j]}^t} \right)\right\}.
\end{align*}

Since $i^{\ast}(\theta^t, x) \le i^{\ast}(\theta^t, z)$, we can show that
\begin{align*}
    &\sum_{j=1}^{i^{\ast}(\theta^t, z)}{\phi_{[j]}^{t}q_{[j]}^t} + \phi_{[i^{\ast}(\theta^t, z) +1]}^t \left(x - \sum_{j=1}^{i^{\ast}(\theta^t, z)}{q_{[j]}^t} \right) - R^t(\theta^t, q^t, x) \\
    & = \sum_{j=i^{\ast}(\theta^t, x)+1}^{i^{\ast}(\theta^t, z)}{\phi_{[j]}^{t}q_{[j]}^t}  \\
    & \qquad + \phi_{[i^{\ast}(\theta^t, z) +1]}^t \left(x - \sum_{j=1}^{i^{\ast}(\theta^t, z)}{q_{[j]}^t} \right) -  \phi_{[i^{\ast}(\theta^t, x) +1]}^t \left(x - \sum_{j=1}^{i^{\ast}(\theta^t, x)}{q_{[j]}^t} \right) \\
    & = \sum_{j=i^{\ast}(\theta^t, x)+1}^{i^{\ast}(\theta^t, z)}{(\phi_{[j]}^{t} - \phi_{[i^{\ast}(\theta^t, z)+1]}^t)q_{[j]}^t} \\
    & \qquad + (\phi_{[i^{\ast}(\theta^t, z) +1]}^t  -  \phi_{[i^{\ast}(\theta^t, x) +1]}^t )\left(x - \sum_{j=1}^{i^{\ast}(\theta^t, x)}{q_{[j]}^t} \right) \\
    & = \sum_{j=i^{\ast}(\theta^t, x)+2}^{i^{\ast}(\theta^t, z)}{(\phi_{[j]}^{t} - \phi_{[i^{\ast}(\theta^t, z)+1]}^t)q_{[j]}^t} \\
    & \qquad + (\phi_{[i^{\ast}(\theta^t, x) +1]}^t  -  \phi_{[i^{\ast}(\theta^t, z) +1]}^t )\left\{ q_{[i^{\ast}(\theta^t, x)+1]}^t - \left(x - \sum_{j=1}^{i^{\ast}(\theta^t, x)}{q_{[j]}^t} \right) \right\} \\
    & \ge 0.
\end{align*}

Similarly, since $i^{\ast}(\theta^t, z) \le i^{\ast}(\theta^t, y)$, 
\begin{align*}
    &\sum_{j=1}^{i^{\ast}(\theta^t, z)}{\phi_{[j]}^{t}q_{[j]}^t} + \phi_{[i^{\ast}(\theta^t, z) +1]}^t \left(y - \sum_{j=1}^{i^{\ast}(\theta^t, z)}{q_{[j]}^t} \right) - R^t(\theta^t, q^t, y) \\
    & = \sum_{j=i^{\ast}(\theta^t, z)+1}^{i^{\ast}(\theta^t, y)}{\phi_{[j]}^{t}q_{[j]}^t} \\
    & \qquad + \phi_{[i^{\ast}(\theta^t, z) +1]}^t \left(y - \sum_{j=1}^{i^{\ast}(\theta^t, z)}{q_{[j]}^t} \right) -  \phi_{[i^{\ast}(\theta^t, y) +1]}^t \left(y - \sum_{j=1}^{i^{\ast}(\theta^t, y)}{q_{[j]}^t} \right) \\
    & = \sum_{j=i^{\ast}(\theta^t, z)+1}^{i^{\ast}(\theta^t, y)}{(\phi_{[j]}^{t} + \phi_{[i^{\ast}(\theta^t, y)+1]}^t)q_{[j]}^t} \\
    & \qquad + (\phi_{[i^{\ast}(\theta^t, z) +1]}^t  -  \phi_{[i^{\ast}(\theta^t, y) +1]}^t )\left(y - \sum_{j=1}^{i^{\ast}(\theta^t, z)}{q_{[j]}^t} \right) \\
    & \ge 0.
\end{align*}

Therefore, $R^t(\theta^t, q^t, z) \ge \alpha R^t(\theta^t, q^t, x) + (1-\alpha)R^t(\theta^t, q^t, y)$, which completes the proof.

\noindent (c) Let $x \le Q$, $y \le R$, and $\alpha \in (0, 1)$ be given. Without loss of generality, suppose $Q < R$. Then,
\begin{align*}
    R^t(\theta^t, \alpha Q + (1 - \alpha)R, \alpha x + (1- \alpha) y) & = R(\theta^t, R, \alpha x + (1- \alpha) y) \\
    & \ge \alpha R(\theta^t, R, x) + ( 1- \alpha) R(\theta^t, R, y) \\
    & = \alpha R(\theta^t, Q, x) + ( 1- \alpha) R(\theta^t, R, y). 
\end{align*}
\end{proof}

\subsection*{Proof of Lemma 3}
\begin{proof}[\unskip\nopunct]

\noindent (a) We will prove the Lemma 3 using mathematical induction. Denote the optimal solution as $x^{\ast}(\theta^t, q^t)$.

At the end of the study period, i.e., $t = T$, let $q^T_1 < q^T_2$, then
\begin{align*}
    V^T(\theta^T, q_1^T) & = R^T(\theta^T, q_1^T, x^{\ast}(\theta^T, q^T_1)) 
     = R^T(\theta^T, q_2^T, x^{\ast}(\theta^T, q^T_1)) \\
    & \le R^T(\theta^T, q_2^T, x^{\ast}(\theta^T, q^T_2)) 
     = V^T(\theta^T, q_2^T).
\end{align*}
The inequality holds because the feasible set of $x$ in $q^T_1$ is included to $q^T_2$.

Then, let's assume that the statement holds in the case of $t+1$. Then, for $q^t_1 < q^t_2$, 
\begin{align*}
    V^{t}(\theta^t, q^t_1) & = R^t(\theta^t, q^t_1, x^{\ast}(\theta^t, q^t_1)) + \delta \mathbb{E}_{n^{t+1}, \theta^{t+1}}[V^{t+1}(\theta^{t+1}, q^t_1 - x^{\ast}(\theta^t, q^t_1))] \\
    & \le R^t(\theta^t, q^t_2, x^{\ast}(\theta^t, q^t_1)) + \delta \mathbb{E}_{n^{t+1}, \theta^{t+1}}[V^{t+1}(\theta^{t+1}, q^t_2 - x^{\ast}(\theta^t, q^t_1))] \\
    & \le R^t(\theta^t, q^t_2, x^{\ast}(\theta^t, q^t_2)) + \delta \mathbb{E}_{n^{t+1}, \theta^{t+1}}[V^{t+1}(\theta^{t+1}, q^t_2 - x^{\ast}(\theta^t, q^t_2))] \\
    & = V^t(\theta^t, q^t_2).
\end{align*}
The first inequality holds by the assumption and the second inequality holds because of the optimality of $x^{\ast}$. 

\noindent (b) Mathematical induction is applied again. For $t = T$, 
\begin{align*}
    V^T(\theta^T, q^T) = R^T(\theta^T, q^T, x^{\ast}(\theta^T, q^T)).
\end{align*}
Without loss of generality, assume that $\phi_i^T > 0$ for every buyer $(i, T)$. Then, $V^T(\theta^T, q^T) = R^T(\theta^T, q^T, q^T)$. Then, from (c) of Lemma 2, $V^T(\theta^T, q^T)$ is concave with respect to $q^T$.

Then, assume that the statement holds in the case of $t+1$. Then, let $q^t_1$, $q^t_2$, and $\alpha \in (0, 1)$ be given. For $x_1 \in [0, q^t_1]$ and $x_2 \in [0, q^t_2]$, define $x$ and $q$ as
\begin{equation*}
    x = \alpha x_1 + (1 - \alpha) x_2 \mbox{ and } q = \alpha q^t_1 + (1-\alpha) q^t_2.
\end{equation*}

Since $V^{t+1}(\theta^t, q^{t+1})$ is concave with respect to $q^{t+1}$,
\begin{align*}
    V^{t+1}(\theta^{t+1}, q - x) &= V^{t+1}(\theta^{t+1}, \alpha (q^t_1 - x_1) + (1 - \alpha)(q^t_2 - x_2)) \\
                        & \ge \alpha V^{t+1}(\theta^{t+1}, q^t_1 - x_1) + ( 1 - \alpha)V^{t+1}(\theta^{t+1}, q^t_2 - x_2).
\end{align*}

It implies that 
\begin{align*}
        \mathbb{E}_{n^{t+1}, \theta^{t+1}}[V^{t+1}(\theta^{t+1}, q - x)] \ge &  \alpha \mathbb{E}_{n^{t+1}, \theta^{t+1}}[V^{t+1}(\theta^{t+1}, q^t_1 - x_1)] \\ 
        & + (1 - \alpha) \mathbb{E}_{n^{t+1}, \theta^{t+1}}[V^{t+1}(\theta^{t+1}, q^t_2 - x_2)].
\end{align*}

Furthermore, since $R^t(\theta^t, q^t, x)$ is concave with respect to $(q^t, x)$, we also have
\begin{align*}
    R^t(\theta^t, q, x) \ge \alpha R^t(\theta^t, q^t_1, x_1) + (1 - \alpha) R^t(\theta^t, q^t_2, x_2).
\end{align*}

Adding up these two inequalities, we have
\begin{align*}
    V^t(\theta^t, q) & = \underset{0 \le x' \le q}{\sup}R(\theta^t, q, x') + \delta \mathbb{E}_{n^{t+1}, \theta^{t+1}}[V^{t+1}(\theta^{t+1}, q - x')] \\
    & \ge R^t(\theta^t, q, x) + \delta \mathbb{E}_{n^{t+1}, \theta^{t+1}}[V^{t+1}(\theta^{t+1}, q - x)] \\
    & \ge \alpha \left\{R^t(\theta^t, q^t_1, x_1) + \delta \mathbb{E}_{n^{t+1}, \theta^{t+1}}[V^{t+1}(\theta^{t+1}, q^t_1 - x_1)] \right\} \\
    & \quad + (1 - \alpha) \left\{ R^t(\theta^t, q^t_2, x_2) + \delta \mathbb{E}_{n^{t+1}, \theta^{t+1}}[V^{t+1}(\theta^{t+1}, q^t_2 - x_2)]\right\}.
\end{align*}
Then, taking the supremum over $x_1 \in [0, q^t_1]$ and $x_2 \in [0, q^t_2]$, we have
\begin{align*}
    V^t(\theta^t, q) \ge \alpha V^t(\theta^t, q^t_1) + (1 - \alpha) V^t(\theta^t, q^t_2),
\end{align*}
which completes the proof.

\noindent (c) Recall that
\begin{equation*}
    \mathbb{E}_{n^t, \theta^t}[V^t(\theta^t, q^t)] = \sum_{n=1}^{N}g(n) \int_{\Theta^N}V^t(\theta^t, q^t)f(\theta^t)d\theta^t.
\end{equation*}
Let $q_1^t < q_2^t$. For given $\theta^t$, since $V^t(\theta^t, q^t_1) \le V^t(\theta^t, q^t_2)$ by (a) of Lemma 3, we have 
\begin{align*}
    \sum_{n=1}^{N}g(n) \int_{\Theta^N}V^t(\theta^t, q^t_1)f(\theta^t)d\theta^t \le \sum_{n=1}^{N}g(n) \int_{\Theta^N}V^t(\theta^t, q^t_2)f(\theta^t)d\theta^t,
\end{align*}
which proves the monotonicity. Furthermore, let $\alpha \in (0,1)$ be given. Then, by (b) of Lemma 3,
\begin{align*}
    V^t(\theta^t, \alpha q^t_1 + (1 - \alpha) q^t_2 ) \ge \alpha V^t(\theta^t, q^t_1) + (1 - \alpha)V^t(\theta^t, q^t_2).
\end{align*}
So, it also follows that
\begin{align*}
    &\sum_{n=1}^{N}g(n) \int_{\Theta^N}V^t(\theta^t, \alpha q^t_1 + (1 - \alpha) q^t_2)f(\theta^t)d\theta^t \\ 
    &\ge \alpha \sum_{n=1}^{N}g(n) \int_{\Theta^N}V^t(\theta^t, q^t_1)f(\theta^t)d\theta^t + (1 - \alpha)\sum_{n=1}^{N}g(n) \int_{\Theta^N}V^t(\theta^t, q^t_2)f(\theta^t)d\theta^t,
\end{align*}
which proves the concavity. 

\end{proof}

\subsection*{Proof of Theorem 5}

\begin{Remark*}
Consider any two type vector $\theta^t$ and $\tilde{\theta}^t$, where $\theta^t = (\theta^t_i, \theta^t_{-i})$ and $\tilde{\theta}^t = (\tilde{\theta}^t_i, \theta^t_{-i})$. Assume that the corresponding virtual value of $\theta^t_i$ is smaller than that of $\tilde{\theta}^t_i$, i.e., $\phi_i^t = v_i^t - \frac{1-F(v^t_i \vert q^t_i)}{f(v^t_i \vert q^t_i)}< \tilde{\phi}_i^t = \tilde{v}_i^t - \frac{1-F(\tilde{v}^t_i \vert \tilde{q}^t_i)}{f(\tilde{v}^t_i \vert \tilde{q}^t_i)}$. Then, the following statement holds.
\begin{equation*}
    x^{\ast}(\theta^t, q^t) \le x^{\ast}(\tilde{\theta}^t, q^t).
\end{equation*}
\end{Remark*}
$\,$
\begin{proof}[Proof of Remark]
$\,$

We first show that $MV^t(\theta^t, q^t, x) \le MV^t(\tilde{\theta}^t, q^t, x)$ for every $x \in [0, q^t]$. Consider the marginal values of state $(\theta^t, q^t)$ and $(\tilde{\theta}^t, q^t)$:
\begin{align*}
    MV^t(\theta^t, q^t, x) &= \frac{\partial}{\partial x} \left\{R^t(\theta^t, q^t, x) + \delta \mathbb{E}_{n^{t+1}, \theta^{t+1}}[V^{t+1}(\theta^{t+1}, q^t - x)] \right\}, \\
    MV^t(\tilde{\theta}^t, q^t, x) &= \frac{\partial}{\partial x} \left\{ R^t(\tilde{\theta}^t, q^t, x) + \delta \mathbb{E}_{n^{t+1}, \theta^{t+1}}[V^{t+1}(\theta^{t+1}, q^t - x)] \right\}.
\end{align*}
Therefore, what we have to show is $\frac{\partial}{\partial x} R^t(\theta^t, q^t, x) \le \frac{\partial}{\partial x} R^t(\tilde{\theta}^t, q^t, x)$. Denote the virtual values of $\theta^t$ as $\phi^t = (\phi^t_1, \cdots, \phi^t_{n^t})$ and the virtual values of $\tilde{\theta}^t$ as $\tilde{\phi}^t = (\tilde{\phi}^t_1, \cdots, \tilde{\phi}^t_{n^t})$. Then, we have $\phi^t_i < \tilde{\phi}^t_i$ and $\phi^t_j = \tilde{\phi}^t_j$ for $(j,t) \neq (i,t)$. 

If $\tilde{\phi}_i^t < \phi^t_{[i^{\ast}(\theta^t, x) +1]}$, the priority of the allocated buyers are unchanged, so
\begin{equation*}
    \frac{\partial}{\partial x} R^t(\theta^t, q^t, x) = \phi^t_{[i^{\ast}(\theta^t, x)+1]} =  \frac{\partial}{\partial x} R^t(\tilde{\theta}^t, q^t, x).
\end{equation*}

If $\tilde{\phi}_i^t \ge \phi^t_{[i^{\ast}(\theta^t, x) +1]}$, there is a chance for the buyer $(i,t)$ to push out the existing winners, and it implies that $\tilde{\phi}^t_{[i^{\ast}(\tilde{\theta}^t, x)+1]} \ge \phi^t_{[i^{\ast}(\theta^t, x) +1]}$. Therefore, 
\begin{equation*}
    \frac{\partial}{\partial x} R^t(\theta^t, q^t, x) = \phi^t_{[i^{\ast}(\theta^t, x)+1]} \le \tilde{\phi}^t_{[i^{\ast}(\tilde{\theta}^t, x)+1]} =  \frac{\partial}{\partial x} R^t(\tilde{\theta}^t, q^t, x),
\end{equation*}
which proves $MV^t(\theta^t, q^t, x) \le MV^t(\tilde{\theta}^t, q^t, x)$ for every $x \in [0, q^t]$. 

Then, by Lemma 5, since both $MV^t(\theta^t, q^t, x)$ and $MV^t(\tilde{\theta}^t, q^t, x)$ is non-increasing with respect to $x$, it is obvious that $X^{-}(\tilde{\theta}^t, q^t) \subseteq X^{-}(\theta^t, q^t)$. Also, if both sets are non-empty, $\inf X^{-}(\theta^t, q^t) \le \inf X^{-}(\tilde{\theta}^t, q^t)$. So, by the definition of $x^{\ast}$, we have $x^{\ast}(\theta^t, q^t) \le x^{\ast}(\tilde{\theta}^t, q^t)$. 

\end{proof}

\begin{proof}[\unskip\nopunct]

\noindent (a) Consider the two types of bidder $(i,t)$: $\theta^t_i = (v_i^t, q_i^t)$ and $\tilde{\theta}^t_i = (\tilde{v}_i^t, q_i^t)$ where $v_i^t < \tilde{v}_i^t$. Under the regularity condition, the corresponding virtual values satisfy $\phi_i^t < \tilde{\phi}_i^t$. Since the virtual values of the other bidders are unchanged, the priority of bidder $(i,t)$ with type $\tilde{\theta}^t_i$ is higher than with type $\theta^t_i$. On the other hand, by the preceding Remark, $x^{\ast}(\theta^t, q^t) \le x^{\ast}(\tilde{\theta}^t, q^t)$. Therefore, we can conclude that $(a_i^t)^{\ast}(\theta^t_i, \theta^t_{-i}, q^t) \le (a_i^t)^{\ast}(\tilde{\theta}^t_i, \theta^t_{-i}, q^t)$. Integrating both sides of the inequality, we obtain $(A_i^t)^{\ast}(\theta^t_i \vert q^t) \leq (A_i^t)^{\ast}(\tilde{\theta}^t_i \vert q^t)$. In other words, the expectations for $\theta^t_{-i}$  still satisfy the inequality.

\noindent (b) The proof of (b) is omitted because it is analogous to (a)

\end{proof}

\subsection*{Proof of Theorem 6}
\begin{proof}[\unskip\nopunct]

\noindent \textbf{Step 1:} \textit{$\Gamma^{\ast}=(a^{\ast}, p^{\ast})$ with the penalty scheme $\rho$ is incentive compatible.}

Let $\theta^t$ and $q^t$ is given. Consider an arbitrary bidder $(i,t)$.  When the bidder has a true type of $\theta^t_i = (v_i^t, q_i^t)$ and reports $\hat{\theta}^t_i = (\hat{v}_i^t, \hat{q}_i^t)$, there are two possibilities of the reported demand: $\hat{q}_i^t \le q_i^t$ or $\hat{q}_i^t > q_i^t$. 

Consider the case where $\hat{q}_i^t \le q_i^t$. Then, the bidder does not have to consider the activation of the penalty scheme, thus the expected utility can be written as 
\begin{align*}
    U_i^t(\hat{\theta}^t_i \vert \theta^t_i) &= \mathbb{E}_{\theta^t_{-i}}[v_i^t a_i^t(\hat{\theta}^t_i, \theta^t_{-i}, q^t) - p_i^t(\hat{\theta}^t_i, \theta^t_{-i}, q^t)] \\
    &= v_i^t A_i^t(\hat{\theta}^t_i \vert q^t) - \left( \hat{v}_i^t A_i^t(\hat{\theta}^t_i \vert q^t) - \int_{0}^{\hat{v}_i^t}A_i^t((u, \hat{q}_i^t) \vert q^t) du \right) \\
    &= (v_i^t - \hat{v}_i^t)A_i^t(\hat{\theta}^t_i \vert q^t) + \int_{0}^{\hat{v}_i^t}A_i^t((u, \hat{q}_i^t) \vert q^t) du. 
\end{align*}
If $v_i^t \le \hat{v}_i^t$, 
\begin{align*}
    &(v_i^t - \hat{v}_i^t)A_i^t(\hat{\theta}^t_i \vert q^t) + \int_{0}^{\hat{v}_i^t}A_i^t((u, \hat{q}_i^t) \vert q^t) du  \\ &= -\int_{v_i^t}^{\hat{v}_i^t} A_i^t((\hat{v}_i^t, \hat{q}_i^t)\vert q^t) du + \int_{0}^{v_i^t}A_i^t((u, \hat{q}_i^t) \vert q^t) du  + \int_{v_i^t}^{\hat{v}_i^t}A_i^t((u, \hat{q}_i^t) \vert q^t) du \\
    & \le \int_{0}^{v_i^t}A_i^t((u, \hat{q}_i^t) \vert q^t) du. 
\end{align*}
The last inequality holds since $A_i^t((v, \hat{q}_i^t)\vert q^t)$ is non-decreasing with respect to $v$. Similarly, if $v_i^t > \hat{v}_i^t$, 
\begin{align*}
    &(v_i^t - \hat{v}_i^t)A_i^t(\hat{\theta}^t_i \vert q^t) + \int_{0}^{\hat{v}_i^t}A_i^t((u, \hat{q}_i^t) \vert q^t) du  \\ &= \int_{\hat{v}_i^t}^{v_i^t} A_i^t((\hat{v}_i^t, \hat{q}_i^t)\vert q^t) du + \int_{0}^{v_i^t}A_i^t((u, \hat{q}_i^t) \vert q^t) du  - \int_{\hat{v}_i^t}^{v_i^t}A_i^t((u, \hat{q}_i^t) \vert q^t) du \\
    & \le \int_{0}^{v_i^t}A_i^t((u, \hat{q}_i^t) \vert q^t) du. 
\end{align*}
Since $\int_{0}^{v_i^t}A_i^t((u, \hat{q}_i^t) \vert q^t) du = U_i^t((v_i^t, \hat{q}_i^t) \vert q^t)$, every bidder has no reason to misreport the marginal value $\hat{v}_i^t \neq v_i^t$. Furthermore, since $A_i^t((v_i^t, q)\vert q^t)$ is non-decreasing with respect to $q$, the bidder's expected utility is maximized when $\hat{q}_i^t = q_i^t$.  

Then, consider the case where $\hat{q}_i^t > q_i^t$. Then, the bidder has to consider the activation of the penalty scheme if the goods are over-allocated. Then, the bidder's expected utility satisfies the following.
\begin{align*}
    U_i^t(\hat{\theta}^t_i \vert \theta^t_i) & =  \mathbb{E}_{\theta^t_{-i}}[v_i^t \min\{q_i^t, a_i^t(\hat{\theta}^t_i, \theta^t_{-i}, q^t)\} - p_i^t(\hat{\theta}^t_i, \theta^t_{-i}, q^t) - \rho_i^t(\hat{\theta}^t_i, \theta^t_{-i}, q^t)] \\
    & \le  \mathbb{E}_{\theta^t_{-i}}[v_i^t a_i^t(\hat{\theta}^t_i, \theta^t_{-i}, q^t) - p_i^t(\hat{\theta}^t_i, \theta^t_{-i}, q^t) - \rho_i^t(\hat{\theta}^t_i, \theta^t_{-i}, q^t)] \\
    & = (v_i^t - \hat{v}_i^t)A_i^t(\hat{\theta}^t_i \vert q^t) + \int_{0}^{\hat{v}_i^t}A_i^t((u, \hat{q}_i^t) \vert q^t) \\
    & \qquad - \int_{(a^t_i)^{\ast, -1}((q_i^t, \infty))} \frac{\bar{v} \hat{q}_i^t}{\mathcal{P}_{\theta^{t}_{-i}}(((a_i^t)^{-1}(\{\hat{q}_i^t\}))} d \mathcal{P}_{\theta^t_{-i}} \\
    & \le (v_i^t - \hat{v}_i^t)A_i^t(\hat{\theta}^t_i \vert q^t) + \int_{0}^{\hat{v}_i^t}A_i^t((u, \hat{q}_i^t) \vert q^t) - \bar{v} \hat{q}_i^t\\    
    & \le 0. 
\end{align*}
The second inequality holds since $\frac{\mathcal{P}_{\theta^{t}_{-i}}((a_i^t)^{-1}((q_i^t, \infty))}{\mathcal{P}_{\theta^{t}_{-i}}(((a_i^t)^{-1}(\{\hat{q}_i^t\}))} \ge 1$, and the last inequality holds since $0 \le A_i^t((v, \hat{q}_i^t) \vert q^t) \le \hat{q}_i^t$ for every $0 \le v \le \bar{v}$. Then, the bidder has no reason to overbid the demand regardless of the reported value $\hat{v}_i^t$, which completes the proof of incentive compatibility. 

$\,$

\noindent \textbf{Step 2:} \textit{$\Gamma^{\ast}=(a^{\ast}, p^{\ast})$ with the penalty scheme $\rho$ is individually rational.}

For any given state $(\theta^t, q^t)$, consider an arbitrary bidder $(i,t)$. Then, for $\theta^t_{-i} \in \Theta^{n^t-1}$, $a_i^t(\theta^t_i, \theta^t_{-i}, q^t) \ge 0$. It follows that $A_i^t(\theta^t_i \vert q^t) \ge 0$. Also, $a_i^t((0,q_i^t), \theta^t_{-i}, q^t) = 0$. Therefore,
\begin{equation*}
    U_i^t(\theta^t_i \vert q^t) = \int_{0}^{v_i^t} A_i^t((u, q_i^t) \vert q^t) du \ge 0.
\end{equation*}

$\,$

\noindent \textbf{Step 3:} \textit{$\Gamma^{\ast}=(a^{\ast}, p^{\ast})$ with the penalty scheme $\rho$ maximizes the seller's expected revenue.}

Consider the case where $T = 2$. Then, by the revelation principle, any mechanism can be replicated by an incentive compatible, feasible direct mechanism. So, consider a direct mechanism $\hat{\Gamma} = ((\hat{a}^1, \hat{p}^1), (\hat{a}^2, \hat{p}^2))$. Since it is incentive compatible, individually rational, and feasible, the expected revenue of the seller who determines to sell $x$ units in the first period is
\begin{align*}
\Pi(\hat{\Gamma}) & = \mathbb{E}_{n^1, \theta^1}\left[\sum_{i=1}^{n^1}\phi_i^1 \hat{a}_i^1(\theta^1, \tilde{Q}) - \sum_{i=1}^{n^1}U_i^1((0, q_i^1) \vert q^t) \right] \\&  + \delta\mathbb{E}_{n^2, \theta^2}\left[\sum_{i=1}^{n^2}\phi_i^2 \hat{a}_i^2(\theta^2, \tilde{Q} - x) - \sum_{i=1}^{n^2}U_i^2((0, q_i^2)\right] 
\end{align*}

Then, for given state $(\theta^1, \tilde{Q})$ and for any $x \in [0, \tilde{Q}]$, we have
\begin{align*}
    &\sum_{i=1}^{n^1}\phi_i^1 \hat{a}_i^1(\theta^1, \tilde{Q}) - \sum_{i=1}^{n^1}U_i^1((0, q_i^1) \vert q^t) \\
    & \qquad +  \delta\mathbb{E}_{n^2, \theta^2}\left[\sum_{i=1}^{n^2}\phi_i^2 \hat{a}_i^2(\theta^2, \tilde{Q} - x) - \sum_{i=1}^{n^2}U_i^2((0, q_i^2)\right] \\
    & \le \sum_{i=1}^{n^1}\phi_i^1 \hat{a}_i^1(\theta^1, \tilde{Q}) + \delta\mathbb{E}_{n^2, \theta^2}\left[\sum_{i=1}^{n^2}\phi_i^2 \hat{a}_i^2(\theta^2, \tilde{Q} - x)\right] \\
    & \le R^1(\theta^1, \tilde{Q}, x) + \delta \mathbb{E}_{n^2, \theta^2}\left[R^2(\theta^2, \tilde{Q} - x, y)\right] \\
    & \le R^1(\theta^1, \tilde{Q}, x) + \delta \mathbb{E}_{n^2, \theta^2}\left[V^2(\theta^2, \tilde{Q} - x)\right] \\
    & \le \underset{0 \le x \le \tilde{Q}}{\sup}R^1(\theta^1, \tilde{Q}, x) + \delta \mathbb{E}_{n^2, \theta^2}\left[V^2(\theta^2, \tilde{Q} - x)\right] 
\end{align*}
Therefore, we can conclude that $\Pi(\hat{\Gamma}) \le \Pi(\Gamma^{\ast})$ when $T = 2$. 

Then, suppose that the statement holds in the case of $T = \tau$. That is, $\Gamma^{\ast} = (a^t, p^t)_{t=1}^{\tau}$ is a revenue-maximizing mechanism when the study period is $\tau$. Claim that $\Gamma^{\ast} = (a^t, p^t)_{t=1}^{\tau + 1}$ is also a revenue-maximizing mechanism when the study period is $\tau + 1$. For this purpose, consider an arbitrary incentive compatible mechanism $\hat{\Gamma} = (\hat{a}^t, \hat{p}^t)_{t=1}^{\tau + 1}$. For $x \in [0, \tilde{Q}]$, define sub-mechanism  $\Gamma^{\ast}_2 = (a^t, p^t)_{t=2}^{\tau + 1}$ and $\hat{\Gamma}_2 = (\hat{a}^t, \hat{p}^t)_{t=2}^{\tau + 1}$. That is, each sub-mechanism starts at $t=2$ with $\tilde{Q} - x$ units of the goods. By the assumption, we have $\Pi(\hat{\Gamma}_2) \le \Pi(\Gamma^{\ast}_2)$. So, when the seller determines to sell $x$ units at state $(\theta^1, \tilde{Q})$, 
\begin{align*}
& \sum_{i=1}^{n^1}\phi_i^1 \hat{a}_i^1(\theta^1, \tilde{Q}) - \sum_{i=1}^{n^1}U_i^1((0, q_i^1) \vert q^t) + \delta \Pi(\hat{\Gamma}_2) \\    
& \le  \sum_{i=1}^{n^1}\phi_i^1 \hat{a}_i^1(\theta^1, \tilde{Q}) + \delta \Pi(\Gamma^{\ast}_2) \\
& \le R^1(\theta^1, \tilde{Q}, x) + \delta \mathbb{E}_{n^2, \theta^2}[V^2(\theta^t, \tilde{Q}-x)] \\
& \le \underset{0 \le x \le \tilde{Q}}{\sup}R^1(\theta^1, \tilde{Q}, x) + \delta \mathbb{E}_{n^2, \theta^2}\left[V^2(\theta^2, \tilde{Q} - x)\right] \
\end{align*}
Therefore, we can conclude that $\Pi(\hat{\Gamma}) \le \Pi(\Gamma^{\ast})$ when $T = \tau +1$, which completes the proof. 
\end{proof}

\newpage 

\subsection*{Allocation algorithm}
\begin{algorithm}[h]
\caption{Allocation rule for given $x^{\ast}(\theta^t, q^t)$}
\label{alg:alloc}
\textbf{Input}: $\theta^t$, $q^t$, $x^{\ast}(\theta^t, q^t)$

\begin{algorithmic}[1] 
\FOR{$1 \le i \le n^t$}
\STATE Compute $\phi_i^t = v_i^t - \frac{1 - F(v_i^t \vert q_i^t)}{f(v_i^t \vert q_i^t)}$
\ENDFOR
\STATE Reorder buyers: $\phi^t_{[1]} > \cdots > \phi^t_{[n^t]} $
\FOR{$1 \le i \le n^t$}
\STATE{$(a_{[i]}^t)^{\ast}(\theta^t, q^t) = \min\{x^{\ast}(\theta^t, q^t), q_{[i]}^t\}$}
\STATE{$x^{\ast}(\theta^t, q^t) \leftarrow x^{\ast}(\theta^t, q^t) - (a_{[i]}^t)^{\ast}(\theta^t, q^t)$}
\ENDFOR
\STATE \textbf{return} $(a_i^t)^{\ast}(\theta^t, q^t)$ for $1 \le i \le n^t$ 
\end{algorithmic}
\end{algorithm}

\subsection*{Payment algorithm}
\begin{algorithm}[h]
\caption{Payment rule for given $x^{\ast}(\theta^t, q^t)$}
\label{alg:payment}
\textbf{Input}: $\theta^t$, $q^t$, $x^{\ast}(\theta^t, q^t)$, $(a^t)^{\ast}(\theta^t, q^t)$
\begin{algorithmic}[1] 
\FOR{$1 \le i \le n^t$}
\IF{$(a_i^t)^{\ast}(\theta^t, q^t) = 0$}
\STATE{$(p_i^t)^{\ast}(\theta^t, q^t) = 0$}
\ELSE
\STATE{Compute $(a^t)^{\ast \ast}(\theta^t, q^t)$, defined as the optimal allocation with $x^{\ast}(\theta^t, q^t) - (a_i^t)^{\ast}(\theta^t, q^t)$ units}
\STATE{$d_{-i} = (a_{-i}^t)^{\ast}(\theta^t, q^t) - (a_{-i}^t)^{\ast \ast}(\theta^t, q^t)$}
\STATE{$(p_i^t)^{\ast}(\theta^t, q^t) = \sum_{j \neq i} \min\{\underline{v}, (\phi_i^t)^{-1}(\phi_j^t(\theta_j^t))\}d_{j}$}
\ENDIF
\ENDFOR
\STATE \textbf{return} $(p_i^t)^{\ast}(\theta^t, q^t)$ for $1 \le i \le n^t$ 
\end{algorithmic}
\end{algorithm}